# Governance of Spreadsheets through Spreadsheet Change Reviews[1]

Miguel A. Ferreira[*], Joost Visser[*,†]

[*]Software Improvement Group, Amsterdam, The Netherlands

[†]Radboud University Nijmegen, The Netherlands

Email: m.ferreira@sig.eu, j.visser@sig.eu

**ABSTRACT**

*We present a pragmatic method for management of risks that arise due to spreadsheet use in large organizations. We combine peer-review, tool-assisted evaluation and other pre-existing approaches into a single organization-wide approach that reduces spreadsheet risk without overly restricting spreadsheet use. The method was developed in the course of several spreadsheet evaluation assignments for a corporate customer. Our method addresses a number of issues pertinent to spreadsheet risks that were raised by the Sarbanes-Oxley act.*

## 1  INTRODUCTION

Spreadsheets are software developed by non-professional programmers. They are used for accounting, project management, resource planning, and many other business applications.

Often, an individual will make the first version of a spreadsheet for his or her own use. When the spreadsheet proves useful, it is shared with others, and may be developed further to become a full-fledged business-critical application.

However, spreadsheets are notorious for their sensitivity to errors [Pank0 2007]. Serious errors in the many calculations that are carried out under the visible surface can emerge. Research shows that around 50% of spreadsheets used contain major errors [Powell 2007]. This may result in irritating failures, such as wrongly counted votes or double counted incomes, but also in major calamities. An example is the collapse of the Jamaican bank system - the result of spreadsheet errors. This could also happen in the City of London [Croll 2009].

Executives must be accountable for the way in which information technology is applied within the organisation. This includes spreadsheets. For example, the provisions of the Sarbanes-Oxley act on end-user programming also apply to spreadsheets [Panko 2005]. But how can executives take responsibility for the numerous spreadsheets that are used within their organization, invisible to higher management and independently of the IT department?

It would be easy to simply ban the use of spreadsheets for supporting operational processes, and to insist that all information technology is purchased, developed and managed through the

[1]This work is funded by the ERDF through the Programme COMPETE and by the Portuguese Government through FCT - Foundation for Science and Technology, project ref. PTDC/EIA-CCO/108613/2008.

IT department. This approach would ensure risk-controlling measures, such as architecture reviews, code inspections, tests and the handling of incidents cannot be evaded. However, such draconian measures would be largely unenforceable. Spreadsheets owe their popularity to their low skill entry level and ease of adjustment. Domain experts can quickly automate all kinds of actions without the intervention of third parties. A ban on spreadsheets would almost certainly make an organization far less effective and flexible.

Restricting the risks of spreadsheets without losing their strength through a ban or stifling control measures is challenging. In this paper, we address this challenge by formulating a pragmatic method for spreadsheet risk management that relies on peer-review for the bulk of spreadsheets and reserves in-depth tool-assisted analysis for a selection of critical spreadsheets. To formulate the method, we used our experience with tool-assisted evaluation assignments of corporate spreadsheets.

In Section2,we introduce a new method called Spreadsheet Change Reviews. We present its main phases and workflow and how it addresses the requirements we define for organization-wide spreadsheet governance. Sections 3 and 4 explain in detail the two key elements of the method, respectively, peer-reviewing and tool-assisted evaluation. Section 5 discusses how Spreadsheet Change Reviews address or enable some of the risk control issues raised by SOX. Finally, Sections 6,7 and 8 present related work, limitations and conclusions, respectively.

## 2  SPREADSHEET CHANGE REVIEWS

In the past years, we carried out several assignments where we evaluated the quality and risks of spreadsheets used in a corporate environment. In each case, we combined tool-based analysis of the spreadsheets themselves with interview-based analysis of the work processes in which the spreadsheets were involved. The results of the evaluations were reported to the spreadsheet users, their direct managers, and the heads of the IT and financial departments.

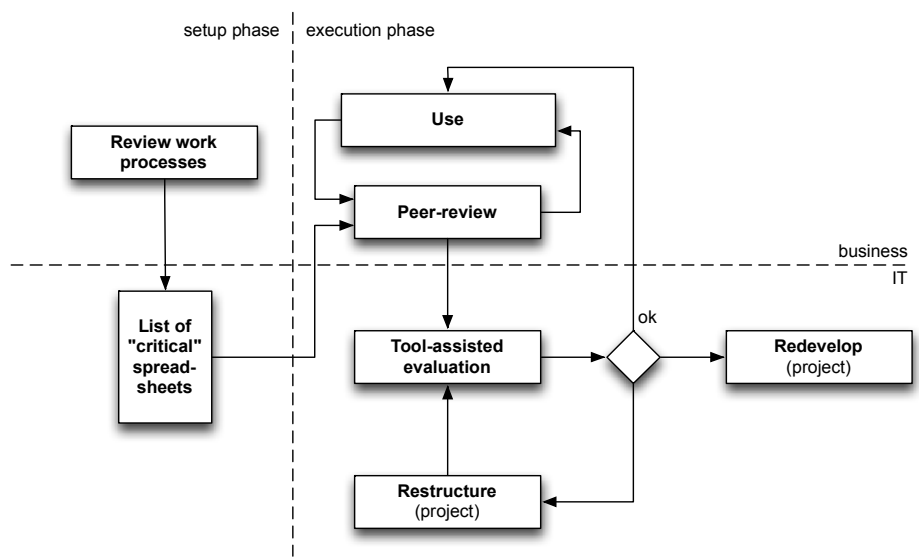

Figure 1: Workflow of Spreadsheet Change Reviews

Based on our experiences in these assignments and the feedback from our customers, we have formulated an approach to spreadsheet governance named Spreadsheet Change Reviews (SCR). The workflow of the SCR is depicted in **Figure 1**.

The core of this approach is to have a lightweight auditing process for every *structural change* [Hunt, 2009] made to a spreadsheet. The process consists of a peer-review where a spread-sheet user asks a designated colleague to review and make a formal statement regarding the changes made to a spreadsheet. If the reviewer finds the change to be correct and to not introduce additional risks, then he formally states that, and the user proceeds with use of the changed spreadsheet. This is depicted in the upper-right quadrant of **Figure 1**.

If, on the contrary, the reviewer does not feel confident about the correctness of the change or the impact it might have, then he will decline to share the responsibility for it. In this case, the user or his manager has the opportunity to ask for a tool-assisted evaluation of the spreadsheet, carried out or contracted by the IT department or quality department. This evaluation leads to recommendations for restructuring, redeveloping, or continuing use of the changed spreadsheets. Any recommendations are discussed with the author of the change, the reviewer, their manager and the executive that decides about the budgetary or strategic implications of these recommendations. This part of the workflow is depicted in the lower-right corner of **Figure 1**.

Rather then entering into these parts of the workflow (the execution phase) directly, an organisation should pass through a *setup* phase. In this phase, work processes are analysed in order to understand how spreadsheets are developed and used in the organisation. An inventory is made of spreadsheets that are in use and these are classified according to their criticality to the business of the organization. The workflow analysis is done in a distributed fashion by representatives of the business (upper-left quadrant in **Figure 1**), while the IT department takes responsibility for merging and managing the results (lower-left quadrant). Additionally, training is needed to prepare users and reviewers for the controls and tools used in the execution phase.

Let us now discuss each of the two phases in more detail.

## 2.1 Setup phase

The setup phase starts with a review of the relevant work processes that the organization has in place. The goal is to understand how spreadsheets are created, modified and used in the context of specific business processes. Information regarding policies, guidelines and best practices used within the organization should also be collected. The purpose behind understanding the way of working of the organization is to be as less intrusive as possible and to take the organization's context into account when defining the controls used in the execution phase. For example, if the organization is already using a version control system that suits their purpose then there is no point in changing that. The same holds true for access control schemes, spreadsheet formatting standards, VBA coding standards, etc.

Because spreadsheets can be used incidentally as throwaway back of the envelope calculations or as key elements in important business processes, it is important to create an inventory of all spreadsheets that are used to support business processes. A spreadsheet classification scheme should also be part of such inventory. The inventory is one of the basic controls that need to be in place according to the SOX recommendations (See Section 5 for details). Previous literature has proposed methods for spreadsheet discovery, triage and classification, as well as for inventory management and maintenance [Chambers, 2008] [Perry, 2008]. Such methods could in principle be re-used in different settings, but it is important to involve the people from the target organization in order to understand what fits them best and to obtain their commitment.

The setup phase terminates with the decision of how the spreadsheet reviews will be performed, what are the mechanisms to request the external evaluation (when necessary), and a training session to prepare the involved stakeholders for the execution phase. This training session should include an overview of most common spreadsheet errors and associated risks, as this has been shown to help decrease error rates [Purser, 2006]. It is worth to mention that there is no one-size-fits-all approach to spreadsheet development and governance [Colver, 2004] [Lemon, 2010]. Therefore, the desired level of structure, formatting and other restrictions should be discussed and agreed upon by all involved stakeholders. Top management, preferably reinforced by their presence in the kick-off and/or training sessions, should communicate the outcomes of the setup phase.

## 2.2 Execution phase

The execution phase is a continuous cycle of use, change and review of spreadsheets (which is considered a good business practice [Hunt, 2009]). In the initial iteration all critical spreadsheets identified in the inventory created in the previous phase should be subject to an in-depth review. The same in-depth review should be conducted whenever a new spreadsheet is added to the inventory (this particular detail is not depicted in **Figure 1**). After the in-depth review, and assuming the reviewer was satisfied with the quality of the spreadsheet, it iscleared for use in the respective business processes.

While in use, every time a structural change is made to a spreadsheet a change review is required. The author of the change requests that one of his peers review the change. For the review, the reviewer uses both a description of the change and a spreadsheet difference detection tool (e.g. an implementation of the methodology proposed in [Hunt, 2009]) to narrow the areas of the spreadsheet that need to be checked. Once the spreadsheet has been reviewed, the reviewer needs to decide whether the change is acceptable or it degraded the quality of the spreadsheet, which can ultimately increase its riskiness. If, on the one hand, the reviewer considers the change to be acceptable, then the reviewer produces a formal approval statement assuming a share of the responsibility for the quality and impact of the change. In this case the spreadsheet is cleared for further use in the respective business processes. If the reviewer does not feel confident in approving the change he requests a tool-assisted spreadsheet evaluation. The change reviews are discussed in greater detail and with examples in Section 3.

There are three possible outcomes from the tool-assisted evaluation: 1) the change is approved, 2) the change is not approved and the spreadsheet should be improved and re-evaluated, or 3) the change is not approved and the issues found are so concerning that the spreadsheet should be re-developed in order to accommodate the necessary changes.

## 2.3 Discussion

The method outline above is designed to satisfy a number of important requirements that we will discuss in turn.

**Cater for the lack of centralized ownership of spreadsheets and lack of explicit budgeting for spreadsheet activities.**

On the one hand, the executives that have the power and responsibility to enforce spreadsheet governance, most of the times, do not handle the critical spreadsheets in their daily activities. On the other hand, the end-users that do handle the critical spreadsheets daily do not have the power or the budget to enforce spreadsheet governance. This dichotomy has a tremendous impact on the implementation of any spreadsheet governance approach. The approach proposed in this paper, addresses this issue by packaging a solution that can be enforced and paid for by whom is running the risks of failure due to spreadsheet errors, while leaving the spreadsheet specific decisions (e.g. to review, to request evaluation, to approve, to disapprove, to restructure, to redevelop) to the people that are closer to the problems and handle these spreadsheets more frequently.

**Create increased auditability and traceability for spreadsheets**

This requirement stems from the regulations that have been put forward for financial reporting and end user computing. In order to increase auditability and traceability of spreadsheets it is necessary to record meta-information about these spreadsheets and their evolution. By installing a process for managing spreadsheet risks that requires controls over structural changes and by documenting the steps of this process, one is already working towards creating an audit trail. Another important aspect in the involvement of peers is the spreading of knowledge about the critical spreadsheets among different people within the organization. The knowledge that is shared is both the contextual knowledge about how and why the spreadsheets evolved and the knowledge that is encoded in each of the reviewed spreadsheets. This does not only apply to financial reporting. In any other domain, organizations still need to learn from mistakes in order to prevent them from reoccurring. If it is hard to find a problem and to trace its origin, then it will also be hard to understand why it was introduced. Without understanding the origin of a problem the chances of preventing it in the future are slimmer.

**Take into consideration the context of the target organization and do not ever-restrict spreadsheet users.**

This requirement is related to what is the right level of control is for the target organization. For instance, it is easy to understand that different levels of control are required for the same company depending on whether that company is publicly traded or not. Thus, finding the right amount of control is paramount to the success of the approach. If there is not enough control, the organization will be incurring in unnecessary risks.Otherwise, if the controls are too strict, there is a high likelihood that the spreadsheet users stop enforcing them altogether.

## 3   SPREADSHEET PEER-REVIEW

Spreadsheet peer-reviews are a critical element in spreadsheet governance [PricewaterhouseCoopers, 2004] [Coster, 2011] [Murphy, 2006] [O'Beirne, 2010]. We identify the following benefits from involving peers in the review process:

- Dissemination of knowledge about the organization's spreadsheets by multiple people.

- Dissemination of spreadsheet development knowledge throughout the organization (since spreadsheet users are most often also spreadsheet developers [Purser, 2006]).

- Raising overall awareness of spreadsheet quality issues in the organization.

- Shared responsibility for spreadsheet quality controls and risk management.

We propose a lightweight method for peer reviewing based on checklists developed together with the people from the target organization. The checklists should cover all the controls defined by the target organization as required for their particular case.

The spreadsheets should be reviewed when they are added to the inventory and when, after being part of the inventory, they undergo a structural change. When a spreadsheet is added to the inventory, the review should be complete and in-depth. When a spreadsheet is changed the review should focus only on the changed parts and the parts directly connected to those. This means that there should be two different checklists, one for each of the enumerated purposes.

Organizations should consider both generic and specific controls. Generic controls are applicable to all spreadsheets, whereas specific controls depend on the domain and purpose of the spreadsheet. Examples of generic controls are (see also [PricewaterhouseCoopers, 2004]):

- Version management: store each version of the spreadsheet;

- Change management: record changes made to the spreadsheet, including author, change description and date of change;

- Assess restrictions: restrict access to the spreadsheet to the people that require it and define different levels of access (e.g. viewing or editing);

- Input restrictions: restrict the users' ability to edit spreadsheet cells to avoid that logic, or constant data is overwritten;

- Backup procedures: define and implement regular backups of the spreadsheet taking in consideration the frequency of use and change;

- Archiving procedures: define and implement archiving procedures to store the versions of the spreadsheet and included data that might be required in audits;

- Separation of concerns: keep inputs, computations and outputs separate for increased maintainability;

- Expected values: define expectations for certain computed values (e.g.: percentages are between 0 and 100, totals add up, or certain values are never negative).

With respect to the specific controls it is necessary to consider a particular spreadsheet, or a particular domain of application. If, for instance, the target organization is dealing with spreadsheets for financial reporting it might decide to adopt controls such as the ones described in [Colver, 2008].

The outcome of the peer-review is either the approval or disapproval of the spreadsheet. If the reviewer decides to approve the spreadsheet he or she confirms that the author of the change took the necessary measures to mitigate risks, thus sharing the responsibility for the status of the spreadsheet. If the reviewer decides not to approve the spreadsheet, it is then subject to tool-assisted analysis by an external and independent party.

In case of approval of the spreadsheet change the reviewer shall produce a statement as follows:

*(… list of changes in the spreadsheet …)*

*I attest to have reviewed the spreadsheet changes listed above against the defined spreadsheet controls and found no nonconformities.*

*To the bets of my knowledge the adoption of these changes does not introduce additional operational risk.*

## 4 TOOL ASSISTED SPREADSHEET EVALUATION

Whenever a peer-review results in the disapproval of a spreadsheet, a tool-assisted analysis by an external and independent party is requested. Although there are several approaches for spreadsheet evaluation, we argue that in order to provide a quick and actionable diagnosis of the spreadsheet a metric-based automated analysis fits the purpose best.

Previous research [Correia, 2011] demonstrated that metrics for spreadsheets follow a power law-like distribution, as do metrics for software. The Software Improvement Group (SIG) has for many years employed metrics for diagnosing problems in software systems. Having pinpointed problematic areas, root cause analysis uncovers the underlying problems that need to be addressed to minimize risks of failure. It is only when the underlying problems are identified that actionable recommendations can be provided. This is to say that blind optimization of bad metrics scores, although very actionable, has less added value to the target organization. We argue that the same approach is applicable to spreadsheets. In fact, we have applied this approach with success to three large spreadsheets of SIG's customers.

Spreadsheet metrics allow the identification and quantification of the magnitude of several bad-practices in spreadsheet development (e.g. long formulas and constants used in formulas). Measures for size can be expressed in number of cells, columns, rows, sheets, formulas, unique formulas and data elements. Other metrics exist also for coupling between formulas and the referenced cells, number of inconsistent cells (either because there is data mixed with formulas, or different formulas mixed with each other, in the same rows or columns). Moreover, cell types and cell block orientation can also be detected through metrics. For example, by counting the number of formulas that are not referenced elsewhere in a spreadsheet one can know how many computation endpoints exist (outputs); by comparing the number of rows to the number of columns in a sheet or cell block one can understand what is its special orientation. More examples of metrics and their use can be found in [Correia, 2011].

As with the peer-review, there are two possible outcomes from the tool-assisted analysis, either the change is approved, or not. If the change is approved, then it is ready for use again. Otherwise, depending on the degree of the problems the external evaluator might recommend the restructuring or the redevelopment of the spreadsheet. In this case, the evaluation report should enumerate all the issues found. If the recommendation is that the spreadsheet is restructured and re-evaluated, then the evaluation report should also enumerate the areas of the spreadsheet that need to be improved. If the recommendation is that the spreadsheet is redeveloped, the evaluation report should explain why is it that the actual structure of the spreadsheet inhibiting the necessary change and how could it be restructured in such a way that it caters for the change while not degrading the quality of the spreadsheet. After the redevelopment, the new spreadsheet is added to the inventory which triggers the procedure for new additions already explained in Section 3.

## 5   RELATION TO SARBANES-OXLEY ACT

The Sarbanes-Oxley act (SOX) [U.S. Government, 2002] defines enhanced controls for auditability and accountability in financial reports. This also includes controls over end-user-computing platforms such as spreadsheets. Section 404 of SOX defines requirements for increasing auditability and traceability of financial reporting, including the spreadsheets that are used in the processes. Internal controls for completeness and correction are a big part of such requirements. Furthermore, the companies are required to present evidence of both self conducted assessments and external audits on the adequacy and effectiveness of their internal controls over financial reporting. Although our objective is not to discuss SOX's Section 404 in detail, because others have already done that to great extent [Panko, 2005], let us discuss some relevant contact points between the recommendations of Section 404 and the approach for spreadsheet governance presented in this paper.

**Responsibility for internal controls** – According to SOX, the responsibility for the effective implementation of internal controls is shared among all executives of an organization and is typically lead by the CFO. However, the executives do not even see all the critical spreadsheets in their departments. Therefore, it stands to reason that more people are involved in the process of implementing and operating the internal controls. Our approach caters for both the attribution of responsibility to executives, as stated in SOX, and for the involvement of the people that develop and operate the spreadsheets on a regular base. On the one hand, the executives are responsible for implementing the SCRs, they have the budget and the power to enforce it. On the other hand, the people that know the spreadsheets best have a key role in the reviewing process, as they are asked to either approve a spreadsheet sharing the responsibility for its quality, or to require an external evaluation.

**External evaluators/auditors** – SOX requires both internal and external audits of critical spreadsheets. Although the approach proposed here is not a full-blown audit, there is the option for having spreadsheets evaluated by an external and independent party. We propose that such evaluation can be supported with automated measurements of spreadsheets. However, other approaches exist, for instance having the spreadsheet rebuilt by a specialized professional according to some standard methodology and compare the outcomes.

**Spreadsheet inventory** – One of the first measures required by SOX is the creation of an inventory of spreadsheets that are relevant for the financial reporting processes. Also all the spreadsheets found should be classified according to their role and criticality. In the setup phase, we propose to create such an inventory, only not restricted to financial reporting spreadsheets. All business critical spreadsheets should be added to the inventory and classified. Following the classification, an adequate set of controls is defined for each spreadsheet.

**Detective controls should be embedded in the spreadsheets** – The purpose of detective controls is to detect problems as they are introduced. Oursetup phase meets this need by defining, together with the target organization, the appropriate controls for each critical spreadsheet.

**Preventive and corrective controls** – Preventive controls attempt to avoid problems from happening in the first place. Whereas, corrective controls are meant as a remediation to an existing problem. Our approach implements preventive controls in the setup phase, where each critical spreadsheet is reviewed in light of its purpose withina business process. The same controls then become corrective when used *post-mortem* during the execution phase.

**Internal audit committees** – SOX requires organizations to create internal audit committees that are independent from senior management. This recommendation is not directly addressed in the SCR approach. However, the people that are involved in the actual reviewing process will acquire knowledge and skills that will make them good candidates for such a committee.

**Record provisions**– Another requirement from SOX is that companies keep records of all elements involved in their financial reporting, allowing auditors to trace back decisions to the time they were made. One of the generic controls we propose is a backup plan for keeping track of different versions of each critical spreadsheet. One way to embed the record provisions recommendation is to schedule specific backup procedures that are synchronized with the financial reporting cycles. This way, the spreadsheets used for the creation of financial reports, including the actual data, can be backed up and retrieved when needed.

# 6 RELATED WORK

There are several contacts points between the method proposed in this paper and the literature on spreadsheet regulations, governance, controls, analysis, tooling, etc. We identify two tracks of literature as the most relevant to discuss here: 1) Organizational approaches to spreadsheet governance; 2) Risk control elements that integrate cleanly in the method we propose.

From the literature on organizational approaches to spreadsheet governance, we discuss the approaches of [Chambers, 2008] [Lemon, 2010]. In both cases the authors identify spreadsheet risk mitigation as their main goal, but Chambers goes one step further in pointing out that the solution should not hinder the productivity of spreadsheet users. We share the same vision of controlling spreadsheet risks while maintaining high productivity. **Table 1** compares the different practical solutions to spreadsheet governance proposed in this paper and by Chambers and Lemon. The most relevant differences are in the fact that in the SCR most items have to be agreed upon with the target organization to achieve the best balance between control, flexibility, productivity and user experience. In contrast, both Chambers and Lemon enumerate specific elements that were employed in their case studies. We find the specific items enumerated in the two papers to be quite relevant. However in order for the governance approach to be useful to, and sustainable at, an organization there must be a full commitment from the relevant people in that organization. By discussing and jointly agreeing on what are the specific elements to put in place we expect to achieve a greater commitment than what would be possible by independently defining those elements. One important issue raised in [Chambers, 2008] is the difficulty in aligning departmental initiatives with organization-wide aspects. We address this issue by providing the same organization-wide framework for spreadsheet governance while giving different departments (or even groups within departments) the freedom to define what is the appropriate level of control for them. Another relevant issue they raise is spreadsheet ownership, which we also cater for by addressing management responsibility hand-in-hand with the distributed ownership of spreadsheets.

Other approaches for spreadsheet risk mitigation exist and we selected one of those to exemplify some of the issues we intentionally avoided when designing SCR. We found the approach of [Jafry, 2006] to be technically advanced and sound. They propose an enterprise solution that "hides" the spreadsheets from the users, leaving the manipulation of the spreadsheets to a restrict group of "super users". Normal users would access the spreadsheets indirectly via a web platform. We find this approach to be too intrusive for an organization because it requires a significant amount of infrastructure to be put in place. Furthermore, it imposes a specific workflow for the entire enterprise, whereas our understanding of today's organizations clearly indicates that different departments or groups of users make very different use of spreadsheets. Furthermore, we believe that spreadsheet users are not inclined to "give-up" their spreadsheet platform (e.g. Excel) in exchange for a more restricted platform.

Regarding risk control elements that can be integrated into the SCRs we identified literature covering creation and management of spreadsheet inventories [Perry, 2008], spreadsheet reviews [Coster, 2011] [Murphy, 2006] [O'Beirne, 2010] and detection of spreadsheet changes (also known as diffing) [Hunt, 2009] [Nash, 2003].

Creating an inventory of spreadsheets is one of the first steps (if not the first) to be taken if one wants to introduce any level of governance. However, creating and maintaining such inventory can be quite a daunting task for large organizations. Perry, in [Perry, 2008], proposes an automated approach to discovery, inventory and risk assessment of spreadsheets. The problem is that manually finding all the spreadsheets within an organization is difficult, but even more difficult is to keep track of all the changes to the found spreadsheets. It is here that specific tooling for crawling network devices and indexing spreadsheets can be of use.

Although spreadsheet analysis tool have came a long way, manual review is still considered a critical control element [Coster, 2011]. Effective spreadsheet reviews need to be well structured in order to guide the reviewer through all the important aspects. We agree with the spreadsheet review stages proposed by Murphy [Murphy, 2006]. We consider that the initial stages of *review scope clarification* and *context understanding* to be fundamental in the SCRs. Otherwise the reviewing task can easily become a daunting task and reviewers' performance will deteriorate. Murphy also advocates the use of spreadsheet metrics to provide the reviewer with a sense of what to expect from a given spreadsheet. The SCRs includes spreadsheet metrics in the tool-assisted analysis, but does not exclude the possibility that the reviewers put in place measurement systems to help them in the peer-reviewing process. Finally, O'Beirne [O'Beirne, 2010] includes reviews in the "Top Ten spreadsheet questions" spreadsheet owners should provide answers to if they want to achieve quality and robustness.

Since most of the reviewing work will be done on spreadsheets that change, it is paramount to the success of the method that these changes can be easily isolated for review. To this end, several approaches have been proposed from which we highlight the ones we consider more promising: DiffXL [Hunt, 2009] and SSScan [Nash, 2003]. The key features of DiffXL are the ability of prioritizing changes based on the risk they introduce and background monitoring of spreadsheets coupled with an alert system. The key features of SSScan are portability across operating systems and the creation of a complete audit trail containing change, timestamp and author. This enables the reviewer to reconstruct the spreadsheet at any given point in time and to replay the changes in order.

## 7    LIMITATIONS OF THE SPREADSHEET CHANGE REVIEWS METHOD

Although the SCRs method has not yet been implemented in its entirety, some limitations have already been identified. Such limitations pertain to budgeting, time pressure, change granularity and flexibility issues.

In order to implement the spreadsheet peer-reviews the target organization must allocate budget for the reviewers. Reviewers need to spend time in training, reviewing, and discussing the outcomes of reviews where new risks have been identified. The lack of resources to allocate for reviews is of course an important limitation to the applicability of the SCRs.

Spreadsheets provide great flexibility in use and adaptation. This accustoms users to adapt their spreadsheets on the fly while using them in operational processes. Implementing the SCRs implies that whenever a structural change is made to an important spreadsheet a peer-review process needs to be carried out before the spreadsheet can be further used. In addition to the peer-review process, whenever new risks are identified, an independent party conducts a tool-assisted evaluation. Although the peer-review process can be much faster than the tool-assisted evaluation, both are time consuming and can block operations from proceeding.

The definition of structural change and the granularity at which it is assessed must be shared by all involved parties, otherwise peer-reviews might be requested for too small changes or, on the other hand, some structural changes might go by without being reviewed.

One of the key aspects of the SCRs is the full flexibility given to the target organization in the selection of the spreadsheet controls to be implemented. This attempts to not over-constrain the spreadsheet users while increasing their commitment to quality assurance. On the one hand, when too few controls are selected a false sense of security is gained. On the other hand, too many controls may lead users to bypass the quality assurance steps of the SCRs.

## 8    CONCLUSIONS

We have presented a pragmatic method for spreadsheet risk management in large organizations. We highlight a number of contributions:

- Our method addresses a number of issues raised by the Sarbanes-Oxley act and can therefore help to achieve compliance with governance requirements.

- Our method makes use of the learning capacity of organizations and allows gradual adoption, because it relies on peer-review for the bulk of spreadsheets in use and reserves more involved tool-based evaluation for a selection of critical spreadsheets.

- Our method avoids imposing excessive restrictions on spreadsheet users, which would make organizations less effective by making advantages of spreadsheets inaccessible.

The method was conceived based on our experiences with tool-based analysis of corporate spreadsheets. We envision that the method will evolve further with use. Some future work items include:

- Validate the method by applying it in a realistic setting.

- Further refinement of peer-reviewing checklists. In particular, we expect a number of specific controls to be designed in organization-specific contexts that can then be generalized for common use.

- Standardization of the evaluation process and reporting format in case of tool-assisted evaluation. We have already achieved in reasonable level of standardization in the course of the various analyses carried out so far, but further standardization is likely to be achievable.

| Element | SCR (this paper) | EUC Control Framework [Lemon, 2010] | Controlling End User Computing Applications [Chambers, 2008] |
|---|---|---|---|
| **Inventory & Classification** | Covered | Just inventory | Covered |
| **Spreadsheet design rules** | Covered (specific items agreed upon with the target organization) | • Documentation<br>• Calculation visibility<br>• Labelling<br>• Separation of inputs, calculations and outputs<br>• Cell locking | Not covered/explicit |
| **Controls** | Covered (specific items agreed upon with the target organization) | • Input, calculation and output checking<br>• Change management<br>• File access protection<br>• Archiving | • Version control<br>• Change control<br>• Access control<br>• Business recovery<br>• Documentation<br>• Testing<br>• Tool access restrictions |
| **Spreadsheet Review** | • Specific items agreed upon with the target organization<br>• Triggered by users or automated change detection process<br>• Explicit possible outcomes and follow-up actions | • Design<br>• Integrity<br>• Control requirements<br>• Triggered by automated change detection process | Not covered/explicit |
| **Remediation & Transition** | Not covered | Covered | Not covered/explicit |
| **Training & Awareness** | • Awareness of spreadsheet errors and risks<br>• Training on how to employ the method | • Awareness of spreadsheet errors and risks<br>• Training on control requirements | • Awareness of spreadsheet errors and risks<br>• Training on control requirements<br>• Excel training |
| **Governance & Sustainability** | Not explicit, but promoted by the lightweight nature of the method, and the inclusion of the target organization in selection design, rules, controls, etc. | Study and deployment of most appropriate techniques to support the method (e.g. inventory management, enabling technologies) | Not covered/explicit |
| **Enabling technologies** | • Tool based analysis<br>• Diffing tool<br>• Other tools that the organization finds relevant to employ | ClusterSeven Enterprise Spreadsheet Manager | Not covered/explicit |
| **Metrics** | Spreadsheet specific metrics used in tool based analysis (e.g. formula coupling) | Not covered/explicit | Operations specific metrics (e.g. number of critical end-user applications) |

**Table 1: Comparison of organizational spreadsheet governance elements**